\documentclass[]{article}

\usepackage{epsfig}
\usepackage{graphicx}
\usepackage{psfrag}
\usepackage{subfigure}

\newcommand{\nuph}[3]{Nucl. Phys. \textbf{#1} ({#2}) {#3}}

\newcommand{\prd}[3]{Phys. Rev. \textbf{D{#1}} ({#2}) {#3}}

\newcommand{\prl}[3]{Phys. Rev. Lett. \textbf{{#1}} ({#2}) {#3}}
\newcommand{\plb}[3]{Phys. Lett. \textbf{B{#1}} ({#2}) {#3}}
\newcommand{\prep}[3]{Phys. Rep. \textbf{{#1}} ({#2}) {#3}}

\title{Charmonium from Classical Pure SU(3) Yang-Mills Configurations}
\author{O. Oliveira\thanks{\textit{email}: orlando@teor.fis.uc.pt},
        R. A. Coimbra\thanks{\textit{email}: rita@teor.fis.uc.pt} \\
        Centro de F\'{\i}sica Computacional, Departamento de F\'{\i}sica, \\
        Universidade de Coimbra, 3004-516 Coimbra, Portugal}
\begin{document}

\maketitle

\begin{abstract}
A generalized Faddeev-Niemi ansatz for the gluon field is discussed. In its
simplest parametrization, the ansatz allows a solution of the classical SU(3)
Yang-Mills equations. From these solutions a confining potential for heavy 
quarkonia is defined. The investigation of charmonium spectra  proves
that the potential is able to reproduce the experimental spectra at the level 
of 3\%. Moreover, charmonium leptonic and strong decays are investigated. The
results are in fair agreement with experimental figures and are in line 
with other quark model calculations.
\end{abstract}

\section{Introduction and motivation}

Quantum Chromodynamics (QCD) is the theory describing the interaction between 
quarks and gluons. QCD being a non-abelian gauge theory, besides the 
quark-gluon interaction includes gluon-gluon interactions. It is expected that
these interactions are responsable for the usual baryons and mesons together 
with exotic states like, for example, glueballs and gluelumps.

In QCD, the basic building blocks are the quark and gluon fields. However, it 
appears that nature has a preference for baryons and meson states, i.e. colour 
singlet states, rather than single isolated quarks or exotic states. 
Quarks seems to be permantly confined within hadrons. The theory provides 
hints about the quark and gluon confinement mechanism, but a fully 
explanation is still lacking.

The simplest picture suggesting quark confinement is the singlet potential
computed from Wilson loops \cite{Wil74,Bal01,BaSc93,Nec03}. This potential
can be viewed as the energy between two static colour charges. For large quark
separations, the potential grows linearly making it impossible to separate two
quarks. In this sense, the singlet potential is a confinement mechanism 
applicable to heavy quarks. For light sector, even the idea of describing the 
quark-quark interaction via a non-relativistic potential is questionable. 

Besides the Wilson singlet potential, two popular confinement mechanisms are 
the dual Meissner effect \cite{Na74,Ho76,Ma76,SuIs05} and the vortex 
condensation picture \cite{Ho78,Co79,NiOl79,De97,QuRe05}. In both pictures, 
confinement is explained invoking special types of gluon configurations, 
namely the condensation of magnetic monopoles in the dual picture and the 
percolation of colour vortices in the condensation scenario. Although, a 
considerable amount of work has been done in favour of both mechanisms, their 
precise confirmation is still missing.

The above described scenarios do not exhaust possible explanations for quark
confinement. Many other mechanisms can be found in the literature. Typically,
these mechanisms look for certain types of gluon configuration which, in 
principle, explain why we are not able to detect a free quark. Implicitly, 
they assume that the origin of confinement is within the pure gauge sector. 
The success of the quenched approximation in lattice QCD further supports such
an idea. 

The interaction between heavy quarks, i.e. charm and bottom, can be studied 
with non-relativistic quark models \cite{Do94}. Hopefully, the interquark
potential \cite{Ri79,BuTy81,Ma80,QuRo,Ei80} should be derived from QCD. 
In practice, the potentials are either phenomenological or, at best, 
QCD-motivated. The idea of defining an interquark potential is appealing and, 
if one could solve the theory, the potential should be related to a given 
gluon configuration as the Coulomb potential in QED. However, so far, the 
known solutions of the classical field SU(3) gauge theory, i.e. instantons, do
not suggest quark or gluon confinement.

In this paper we construct solutions of the classical field SU(3) Yang-Mills
theory, in Minkowsky spacetime, which lead to confining nonrelativistic 
potentials\footnote{See \cite{OrCo03,Or05} for related work.}. The classical
solutions are obtained after the introduction of a generalized 
Cho-Faddeev-Niemi-Shabanov ansatz which, in Landau gauge, allows a 
parametrization of the gluon field in terms of two vector fields and a scalar
field. For such a gluonic field, the classical field equations are essentially
a system of coupled QED like equations for the two vector fields and a 
massless Klein-Gordon equation for the scalar field. The solutions of the 
coupled equations can be divided into QED like solutions, i.e. plane wave
solutions, and exponential growing fields. The later ones will allow to 
identify a confining potential for heavy quark systems. Although the potential
is given by a multipole expansion, in this work only the $l=0$ term will be 
considered. In order to compute its parameters, the potential is compared with
the singlet potential computed from the Wilson loop. It comes that, in the 
range 0.2 - 1 fm, the new potential follows closely the Wilson loop behaviour,
with a maximum deviation of about 50 MeV. The differences being both the small 
($ \mathcal{O} (1/r)$) and large ($\mathcal{O} ( \exp{ (\Lambda r)} / r )$) 
distance behaviours. Moreover, the potential is characterized by a single mass
scale $\Lambda = 228$ MeV, computed together with the other parameters by 
fitting the singlet potential in the region 0.2 - 1 fm. Being close to the
singlet potential, the charmonium is well described by the new potential. 
Moreover, assuming a pure vector interaction one can define a spin dependent
potential which, together with central potential, is able to reproduce
experimental spectra with an error of less than 3\%.
The leptonic and sam hadronic decays of charmonium are also investigated. The
model is is fair agreement with the experimental figures.

The paper is organized as follows, in section 2, for completeness, the 
generalized Cho-Faddeev-Niemi-Shabanov ansatz proposed previously is 
discussed, giving particular attention to its simplest parametrization. In
section 3, a potential for heavy quarks is motivated and compared with the 
singlet potential computed from the Wilson loop. The potential parameters are 
computed fitting the singlet potential in the region 0.2 - 1 fm. 
In section 4, the charmonium spectra is calculated with the new potential,
including the effect of spin-dependent forces. The recently discovered 
charmonium states and their possible quantum numbers assignment are discussed. 
Furthermore, discussion of the hadronic decays and leptonic decays of the 
$1^{--}$ is performed. Finally, we draw the conclusions and plans for future 
work in the last section.

\section{Classical Gluon Fields}

For SU(3) the lagrangian density\footnote{In this work we use the notation
of \cite{Hu92}.} reads
\begin{equation}
 \mathcal{L} \, = \, - \frac{1}{4} \, F^a_{\mu\nu} \, F^{a \, \mu\nu}
\end{equation}
where
\begin{equation}
 F^a_{\mu\nu} \, = \, \partial_\mu A^a_\nu \, - \,
                      \partial_\nu A^a_\mu  \, - \,
                      g f_{abc} A^b_\mu A^c_\nu
 \mbox{ } ,
\end{equation}
and $A^a_\mu$ are the gluon fields. The classical equations of motion,
\begin{equation}
 \partial_\mu F^{\mu\nu} ~ + ~ i g \left[ A_\mu \, , \, F^{\mu\nu} \right] ~
  = ~ 0 ,
\label{QCDclasseq}
\end{equation}
being a set of non-linear partial differential equations are quite difficult
to solve and to compute a solution of (\ref{QCDclasseq}) it is
usual to introduce an ansatz.

Following a procedure suggest in \cite{Cho80}, let us consider a real 
covariant constant field $n^a$. From its definition, it follows
\begin{equation}
 D_\mu n ^a ~ = ~ \partial_\mu n^a \, + \, i g ( F^c )_{ab}  A^c_\mu n^b 
 ~ = ~ 0,
\end{equation}
where $F^a$ are the generators of representation to which $n$ belongs. If one
wants to parametrize the gluon field in terms of $n$, the scalar field should
have, at least, as many color components as $A_\mu$. The adjoint representation
fullfils such condition and it will be assumed that $n$ belongs to this 
representation, then
\begin{eqnarray}
 ( F^b )_{ac} & = & - i \, f_{bac} \, , \\
 D_\mu n ^a   & = & \partial_\mu n^a \, + \,  g f_{abc}  n^b A^c_\mu ~ = ~ 0.
 \label{neq}
\end{eqnarray}
Multiplying the last equation by $n^a$ it follows
\begin{equation}
   \partial_\mu \left( n^a n^a \right) ~ = ~ 0 \, .
\end{equation}
From now on it will be assumed that $n^a n^a = 1$. 

Defining the color projected field
\begin{equation}
   C_\mu ~ = ~ n^a A^a_\mu
\end{equation}
one writes the gluon field as
\begin{equation}
   A^a_\mu ~ = ~ n^a \, C_\mu ~ + ~ X^a_\mu \, ,  \label{AnCX}
\end{equation}
where $X$ is orthogonal to $n$ in the sense
\begin{equation}
   n^a X^a_\mu \, = \, 0.
\end{equation}
In order to establish the gauge transformation properties of the various
fields, we consider infinitesimal gauge transformations,
\begin{eqnarray}
  U(x) & = & e^{-i \omega(x) } ~ = ~ 1 - i \omega(x) \\
  \delta A^a_\mu & = & \frac{1}{g} \partial_\mu \omega^a \, + \,
                       f_{abc} \omega^b A^c_\mu  
                                                    \label{Agauge} \\
  \delta n^a & = &  f_{abc} \, \omega^b \, n^c \, . 
                                                    \label{ngauge}
\end{eqnarray}
For $C_\mu$, the transformation rule looks like a
``projected abelian'' gauge transformation
\begin{equation}
  \delta C_\mu ~ = ~  \frac{1}{g} \, n^a \partial_\mu \omega^a \, .
\end{equation}
For $X^a_\mu$, the transformation property is non-linear in the field
$n$. It follows from (\ref{AnCX}), (\ref{Agauge}) and (\ref{ngauge})
\begin{equation}
 \delta X^a_\mu ~ = ~ \frac{1}{g} \left( \delta^{ab} \, - \,  n^a n^b \right) 
                                   \partial_\mu \omega^b 
                     \, + \, f_{abc} \omega^b X^c_\mu
\end{equation}
and
\begin{equation}
   \delta \left( n^a X^a_\mu \right)  ~ = ~ 0.
\end{equation}

In order to build an usefull ansatz, one can go back to equation (\ref{neq}),
replace the gluon field by the decomposition (\ref{AnCX}), multiply by 
$f_{ade} n^e$ and try to solve for the $X$ field. After some algebra we get
\begin{equation}
 f_{abc} n^b \partial_\mu n^c ~ - ~
 \frac{2}{3} \, g \, X^a_\mu  ~ - ~
 g \, \left( d_{abe} d_{dce} - d_{dbe} d_{ace} \right) X^b_\mu n^c n^d 
 ~  = ~ 0 \, \label{eq17}
\end{equation}
after using the SU(3) relation
\begin{equation}
    f_{abc} f_{dec} ~ = ~ \frac{2}{3} \left( \delta_{ad} \delta_{be} \, - \,
                                             \delta_{ae} \delta_{bd} \right)
                        ~ + ~
                        \left( d_{adc} d_{bec} - d_{bdc} d_{aec} \right) \, ,
\end{equation}
where
\begin{equation}
    d_{abc} ~ = ~ \frac{1}{4} \mbox{Tr} 
      \left( \lambda^a \left\{ \lambda^b \, , \, \lambda^c \right\} \right)
\end{equation}
and $\lambda^a$ are the SU(3) Gell-Mann matrices. Equation (\ref{eq17}) allows
us to write
\begin{equation}
  X^a_\mu ~ = ~ \frac{3}{2 g} f_{abc} n^b \partial_\mu n^c ~ + ~ Y^a_\mu
 \label{ansatz2}
\end{equation}
and
\begin{equation}
    n^a X^a_\mu ~ = ~ n^a Y^a_\mu ~ = ~ 0 \, .
    \label{ortho}
\end{equation}
The first term of $X^a_\mu$ is a generalized Faddeev-Niemi ansatz 
\cite{FaNi99}.
The simplest non-trivial $n$ field with norm one can be parametrized as
\begin{equation}
  n^a ~ = ~ \delta^{a1} \sin \theta \, + \, \delta^{a2} \cos\theta.
\end{equation}
Then, with the above definitions equation (\ref{eq17}) implies the following
relations between the $Y$ fields
\begin{eqnarray}
 & &  Y^2_\mu ~ = ~ Y^1_\mu \, \cot \theta \, , \\
 & &  Y^3_\mu ~ = ~ \frac{1}{2g} \, \partial_\mu \, \theta \, , \\
 & &  Y^4_\mu ~ = ~ Y^5_\mu ~ = ~ Y^6_\mu ~ = ~ Y^7_\mu ~ = ~ 0 \, ,
\end{eqnarray}
which together with the color space orthogonality condition (\ref{ortho})
gives the gluon field
\begin{equation}
   A^a_\mu ~ = ~ n^a \, C_\mu 
                           ~ - ~ \frac{\delta^{a3}}{g} \, \partial_\mu \theta
                           ~ + ~ \delta^{a8} \, B_\mu \, ;
  \label{Asimples}
\end{equation}
in the last equation we defined $B_\mu = Y^8_\mu$. The corresponding gluon
field tensor components are
\begin{equation}
   F^a_{\mu\nu} ~ = ~ n^a \, \mathcal{C}_{\mu\nu} ~ + ~
                      \delta^{a8} \, \mathcal{B}_{\mu\nu} \, ,
\end{equation}
where
\begin{eqnarray}
   \mathcal{C}_{\mu\nu} ~ = ~ \partial_\mu C_\nu \, - \, 
                              \partial_\nu C_\mu \, , &  &
   \mathcal{B}_{\mu\nu} ~ = ~ \partial_\mu B_\nu \, - \, 
                              \partial_\nu B_\mu \, .
\end{eqnarray}
For the above gluonic configuration, the classical action is a functional
of the vector fields $C_\mu$ and $B_\mu$,
\begin{equation}
  \mathcal{L} ~ = ~ - \frac{1}{4} 
                      \left( \mathcal{C}^2 \, + \, \mathcal{B}^2 \right) \, ,
\end{equation}
and the classical equations of motion (\ref{QCDclasseq}) become QED-like
equations
\begin{eqnarray}
   \partial^\mu \mathcal{C}_{\mu\nu} ~ = ~ 0 \, ,  &  &
   \partial^\mu \mathcal{B}_{\mu\nu} ~ = ~ 0 \, . \label{FieldEq}
\end{eqnarray}
Moreover, the hamiltonian density and the spin tensor are given by the sum of 
the contributions of two abelian-like theories associated with $C_\mu$ and 
$B_\mu$ fields. 
Note that the pure gauge theory at the classical level, is independent of 
$\theta$. However, the inclusion of fermions implies a coupling 
$\theta$-fermions. The coupling with $\theta$ is associated with the first
three Gell-Mann matrices and, for configurations with $C_\mu = B_\mu = 0$
requires only $\lambda^3$; no coupling with the third color component and the 
coupling to the first two color components have opposite signs.

The computation of classical solutions of the equations for the gluon 
configuration considered requires gauge fixing. 
For the Landau gauge\footnote{For the Coulomb gauge, the first two equations 
are slightly more complicated but in the third and fourth equation one should
replace $\partial$ by $\nabla$, together with other obvious modifications.},
the full set of equations is
\begin{eqnarray}
 \partial^\mu C_\mu ~ = ~ 0 \, ,           &  & 
                              \partial^\mu B_\mu ~ = ~ 0 \, , \label{Feq1} \\
 \partial^\mu \theta \, C_\mu ~ = ~ 0 \, , &  & 
                              \partial^\mu \partial_\mu \theta ~ = ~ 0  \, .
 \label{Feq2}
\end{eqnarray}
These together with the classical field equations have plane wave solutions 
characterized by a four momenta $k$ such that $k^2 = 0$, the fields $C_\mu$ 
and $B_\mu$ are polarized perpendicularly to $k$,
\begin{eqnarray}
 & &  C_\mu (k,\lambda) ~ =  ~ \epsilon_C (k,\lambda) \, e^{-ikx} \, ,  
                   \hspace{1cm}
                   k^\mu \epsilon_C (k,\lambda) ~ = ~ 0 \, , \\
 & &  B_\mu (k,\lambda) ~ =  ~ \epsilon_B (k,\lambda) \, e^{-ikx} \, ,  
                   \hspace{1cm}
                   k^\mu \epsilon_B (k,\lambda) ~ = ~ 0 \, , \\
 & &  \theta (k) ~ = ~ \theta_0 e^{-ikx} \, ,
\end{eqnarray}
and the third color component being polarized along $k$. Our aim is to 
identify possible solutions of the classical field equations (\ref{Feq1}),
(\ref{Feq2}), (\ref{FieldEq}) which can suggest confining solutions. 
We will not proceed with the discussion of the solutions of the above set
of coupled equations but, instead, discuss a particular class of solutions.

For the pure gauge theory, the condition for a finite action/energy solution 
does not constraint $\theta$. Let us consider the simplest non-trivial 
configuration one could think of, namely $C_\mu = B_\mu = 0$. This particular
configuration has zero energy and the field equations are reduced to a 
massless Klein-Gordon equation for $\theta$ which can be solved by the usual
separation of variables. Indeed, writing 
$\theta (t, \vec{r}) = T(t) V ( \vec{r} ) $, it comes
\begin{equation}
   \frac{T^{\prime\prime} (t)}{ T(t) } ~ = ~ 
   \frac{\nabla^2 V ( \vec{r} )}{ V ( \vec{r} ) } ~ = ~ \Lambda^2 \, ,
\end{equation}
where $\Lambda$ is mass scale which is invariant under rescaling of the gluon 
field. The solutions with $\Lambda^2 < 0$ are the usual free field 
solutions. For $\Lambda = 0$, the gluon field is linear in time and the 
spatial part is the usual solution of the Laplace equation. 
If $\Lambda^2 > 0$,
\begin{equation}
  T(t) ~ = ~ A_T e^{\Lambda t} \, + \, B_T e^{-\Lambda t}
  \label{SolT}
\end{equation}
and writing
\begin{equation}
  V( \vec{r} ) ~ = ~ \sum_{l,m} \, V_l (r) \, Y_{lm} ( \Omega ) \,
  \label{SolR}
\end{equation}
where $Y_{lm} ( \Omega )$ are the spherical harmonics, the functions
$V_l (r)$ can be writen in terms of the modified spherical Bessel functions
of the 1$^{\mbox{st}}$ $I_{l + 1/2} (z)$ and 3$^{\mbox{rd}}$ $K_{l + 1/2} (z)$
kind, where $z = \Lambda r$,
\begin{equation}
  V_l (r) ~ = ~ A_{lm} \frac{I_{l + 1/2} (z)}{\sqrt{z}} \, + \,
                B_{lm} \frac{K_{l + 1/2} (z)}{\sqrt{z}} \, .
  \label{SolR1}
\end{equation}
These solutions have the same scale for the temporal and spatial parts and, 
asymptotically, are real exponential functions. For example, the lowest
multipole solution is
\begin{equation}
    V_0 (r) ~ = ~ A \, \frac{ \sinh ( \Lambda r ) }{r} \, + \,
                  B \, \frac{e^{- \Lambda r}}{r}
   \label{V0}
\end{equation}
and the associated gluon field is
\begin{eqnarray}
   A^3_0 & = & \Lambda \left( e^{\Lambda t} - b_T \, e^{- \Lambda t} \right)
               \, V_0 (r) \, , \label{A30} \\
  \vec{A}^3 & = & - \left( e^{\Lambda t} + b_T \, e^{- \Lambda t} \right)
               \, \nabla V_0 (r) \, . \label{A3i}
\end{eqnarray}

\section{A Non-relativistic Potential for Heavy Quarkonium}

\begin{figure}[t]
   \psfrag{EIXOX}{$\Lambda$ (MeV)}
   \psfrag{EIXOY}{$\chi^2$}
   \centering
   \includegraphics[origin=c,scale=0.3]{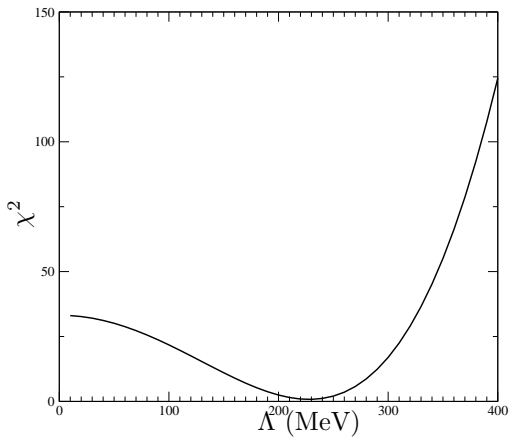}
\caption{Minima of $\chi^2$ as function of $\Lambda$.} \label{Chi2L}
\end{figure}

The coupling of the classical field configuration discussed in the previous 
section to the fermion fields requires only the $\lambda^3$ Gell-Mann matrix, 
i.e.
\begin{equation}
   \gamma^\mu \, \frac{\lambda^3}{2} \, \partial_\mu \theta \, \psi \, ;
\end{equation}
$\psi$ should be understood as a quark Dirac spinor. Assuming that it makes 
sense to describe the heavy fermion interaction via a potential derived from
the above gluon configuration, from the structure of $\lambda^3$ one naively
expects a bound state, let us say the first color component, an unbound state,
the second color component and a free particle solution, the third color 
component. Of course, the usual quark model picture for hadrons is recovered 
only after gauge transforming all fields. 

The classical solution discussed in the last part of the previous section
has the same spatial and time scale $\Lambda = 228 MeV$, i.e. its spatial scale
is $L = 0.865$ fm, while its temporal scale $T = 2.886 \times 10^{-24}$ s.

For charmoninum, typical time scales ranges from  
$\tau = 1/91 \mbox{ KeV}^{-1} = 7.2 \times 10^{-21}$ s for $J / \psi (1S)$ 
up to
$\tau = 8.4 \times 10^{-24}$ s for $\psi (4160)$. 

For bottomonium the time scale ranges from $\tau = 1.5 \times 10^{-21}$ s for 
$\Upsilon (1S)$ up to $\tau = 6.0 \times 10^{-24}$ s for $\Upsilon (10860)$. 

Typical time scales for charmonium and bottomonium are much larger than the 
time scale of the classical configuration. If quarks propagate in the
background of the above classical solution, the effects of propagation should 
be small, i.e. the quark interactions are almost contact interaction in time.
Therefore, in first approximation, the interaction can be parametrized as
\begin{equation}
\delta(t^\prime - t) V( \vec{x}^\prime  - \vec{x}) 
\psi(t^\prime , \vec{x}^\prime ) \psi(t , \vec{x} ) \, .
\end{equation}

On the other end, typical wave lengths of charmonium and bottomonium mesons 
are $\sim 1/3000 \mbox{ MeV}^{-1} = 0.07$ fm and  
$\sim 1/10000 \mbox{ MeV}^{-1} = 0.02$ fm, respectively, and they are much 
smaller than $L \sim 0.9$ fm. From the point of view of the classical 
configuration, the mesons are point like structures, i.e. they can be
described by a point like wave function $\phi ( \vec{r}, t)$. 
Furthermore, if one 
assumes that charmonium and bottomonium are non-relativistic quark systems, 
then it seems reasonable to describe the mesons via a gauged Schr\"odinger 
equation with a non-relativistic potential given by (\ref{SolR}). In first 
approximation one can forget the contribution from the vector potential
of the classical solution.

The non-relativistic potential (\ref{SolR}) is not spherical symmetric. 
However, if only the lowest multipole multipole (\ref{V0}) is considered, one 
is back to the central potential picture. From now on, we will assume that the
confining potential for heavy quarkonium is given by the lowest multipole of 
the spatial part of $A^0_3$ and that the Schr\"odinger equation for heavy 
mesons is, in first approximation,
\begin{equation}
  \left\{ - \, \frac{\nabla^2}{2m} \, + \, V_0 (r) \right\} \phi ~ = ~ 
   i \partial_t \, \phi \,
\end{equation}
where $\phi$ is a heavy meson field and $m$ the reduced mass of the quarkonium
system. 

The potential (\ref{V0}) is coulombic for small quark distantes
\begin{equation}
  V_0 (r) ~ = ~ \frac{B}{r} \, ,
\end{equation}
and confining for large distances
\begin{equation}
  V_0 (r) ~ = ~ \frac{A}{2 r} \, e^{\Lambda r} \, ;
\end{equation}
since the potential grows exponentially with distance, it can be viewed as
a soft bag model. For large inter-quark distances, the wave function goes to 
zero as
\begin{equation}
  \phi ( \vec{r} ) ~ \longrightarrow ~ \frac{1}{r} \,
  \exp \left\{ -\frac{2}{\Lambda} \, \sqrt{ \frac{ \, m \, A}{r} } \,
                \exp  \left[ \frac{\Lambda r}{2} \right] \right\} \, ,
 \hspace{1cm} \mbox{for } r \gg 1 \, ,
\end{equation}
with the spatial extension of the quark field becoming smaller for heavier
quarks.

\begin{figure}[t]
   \psfrag{EIXOY}{$\Delta V$ (MeV)}
   \psfrag{EIXOX}{r (fm)}
   \centering
   \includegraphics[origin=c,scale=0.3]{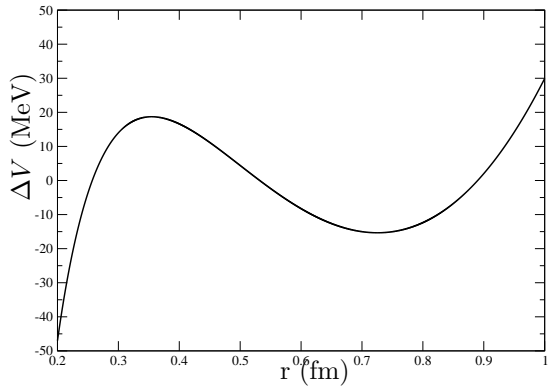}
\caption{$\Delta V = V_0 (r) - V_{singlet} (r)$.} \label{CoimSing}
\end{figure}

The first step towards a proper definition of $V_0$ as a potential for heavy 
quark systems is the computation of its various parameters, namely $A$, $B$ 
and $\Lambda$. This can be done by minimizing the square difference between 
the new potential and the singlet potential computed from Wilson loops. For 
the singlet potential we used the results from \cite{BaSc93} at $\beta = 6.4$.
The
\begin{equation}
   \chi^2 ~ = ~ \int^{r = 1  fm}_{r = 0.2  fm} dr ~
   \bigg[ V_0 ( r ) - V_{singlet} (r) \bigg]^2
\end{equation}
as function of $\Lambda$ can be seen in figure \ref{Chi2L}. The curve shows
an absolute minimum and we take the potential parameters as their values at
this point\footnote{Assuming that the short distance behaviour of the 
inter-quark potential is dominated by the one-gluon exchange contribution, 
then $B = - 4 \alpha_s /3$. The strong coupling constant associated to the 
fitted value for $B$ is $\alpha_s = 0.526$.},
\begin{equation}
  A ~ = ~ 11.2542671, 
  \hspace{0.7cm} B ~ = ~ -0.70113530875, 
  \hspace{0.7cm} \Lambda ~ = ~ 228.026 \, \mbox{MeV}
\end{equation}
when $r$ is measured in MeV$^{-1}$. The difference between $V_0$ and the 
lattice potential is reported in figure \ref{CoimSing}. 

It is curious that the value of the only energy scale in the solution is close
to standard values for $\Lambda_{QCD}$. We have no interpretation for this 
result. Probably, it is connected to the way the parameters are defined, i.e. 
to the fit of $V_0$ to the lattice singlet potential. 

\section{Charmonium - the spectra}

The potential $V_0 (r)$ is a first approximation to the full interaction. From
the various possible corrections, it will be assumed that the spin-dependent
part gives the dominant contribution. Assuming a pure quark-quark vector 
interaction, the spin potential (see, for example, \cite{Grom91} or 
\cite{Do94}) is given by 
\begin{equation}
 V_{spin} ~ = ~  V_{SO} + V_{ten} + V_{SS} ,
\label{Vspin}
\end{equation}
where the spin-orbit $V_{SO}$, tensor $V_{ten}$, and spin-spin $V_{SS}$ 
potentials are
\begin{eqnarray}
 &  &
V_{SO} ~ = ~ 
    \frac{3}{2 m_Q^2} ~ \frac{1}{r} ~ \frac{dV_0}{dr} ~ 
             \vec{L} \cdot \vec{S} \, , \\
 & &
V_{ten} ~ = ~
    \frac{1}{12 m_Q^2} ~ 
    \left(\frac{1}{r}\frac{dV_0}{dr} \, - \, \frac{d^2 V_0}{dr^2} \right)
    ~ S_{ten} \, , \\
 & &
V_{SS} ~ = ~
    \frac{2}{3 m_Q^2} ~ \vec{\nabla}^2 V_0 ~ \vec{s}_q \cdot \vec{s}_{\bar q}
    \, ,
\end{eqnarray}
where $\vec{L}$ is the orbital angular momentum, $\vec{s}_Q$ the spin of the 
heavy quark, $m_Q$ its mass and
\begin{equation}
   S_{ten} ~ = ~ 4 \bigg[ \, 3 \left( \vec{s}_Q \cdot \hat{r} \right)
                              \left( \vec{s}_{\bar Q} \cdot \hat{r} \right) 
                         ~ - ~
                         \vec{s}_Q \cdot \vec{s}_{\bar Q} \, \bigg]
\end{equation}
is the tensor operator. With the above definitions, the potential verifies the
Gromes constraint \cite{Gro84} that arises from the boost invariance of the 
QCD.

In the following, the mesons states are classified according to the usual
spectroscopic notation $n \, ^{2S+1}L_J$, where $S$ is the total spin and $J$ 
the total angular momenta. The Schr\"odinger equation is solved for $V_0 (r)$
with the wave function 
$\Psi_{nlm} ( \vec{r} ) \, = \, R_{nl} (r) \, Y_{lm} ( \theta , \phi )$, where
$n$ is the principal quantum number, $l$ and $m$ are the orbital angular
momentum and its projection, $R_{nl} (r)$ is the radial wave function and
$Y_{lm} ( \theta , \phi )$ are the spherical harmonic functions. The total 
wave function is then built by coupling $\Psi_{nlm} ( \vec{r} )$ with the spin
wave function. The contribution of the spin-dependent potentials will be 
included perturbatively.

We further define
\begin{equation}
   R^{(l)}_{nl} ( 0 ) ~ = ~ \frac{d^l R_{nl} (0)}{dr^l} \, , \label{R0}
\end{equation}
i.e. the radial wave function and its derivatives at the origin. The various
$R^{(l)}_{nl} ( 0 )$ are required to compute production rates and decays rates.

\subsection{Quark mass}

The charm quark mass, $m_c$, was tunned to reproduce the 
$\chi_{c2}(1P) - J/\psi(1S)$ mass difference. Using the mesons mass values
from the particle data book \cite{PDG}, the mass diference is
\begin{equation}
  M \left( \chi_{c2} (1P) \frac{}{} \right) \, - \, 
  M \left( J/\Psi (1S) \frac{}{}\right) ~ = ~
  459.34 \pm 0.11 \, \mbox{MeV} \label{mc}
\end{equation}
and $m_c = 1870$ MeV. We have studied other possible definitions for $m_c$ but
although $m_c$ changes, the meson spectra is essentially the same discussed 
here.

\begin{table}[t]
\begin{center}
\begin{tabular}{cccccc}
\hline\hline
      &\multicolumn{5}{c}{$|R^{(l)}_{nl}(0)|^2$ }\\
      &  \begin{small} $V_0$       \end{small}   
      &  \begin{small} QCD(BT)     \end{small}
      &  \begin{small} Power law   \end{small}
      &  \begin{small} Logarithmic \end{small}
      &  \begin{small} Cornell     \end{small}    \\
\hline\hline
      1S & 1.739 & 0.810 & 0.999 & 0.815 & 1.454 \\
      2S & 1.261 & 0.529 & 0.559 & 0.418 & 0.927 \\
      3S & 1.283 & 0.455 & 0.410 & 0.286 & 0.791 \\
         &       &       &       &       &       \\
      2P & 0.248 & 0.075 & 0.125 & 0.078 & 0.131 \\
      3P & 0.394 & 0.102 & 0.131 & 0.076 & 0.186 \\
         &       &       &       &       &       \\
      3D & 0.222 & 0.015 & 0.026 & 0.012 & 0.031 \\
\hline\hline
\end{tabular}
\caption{Charmonium radial wave function and its derivatives at the origin.
All quantities are in GeV$^{2l+3}$. The first column reports the value 
computed using $V_0 (r)$ and all others are taken from \cite{EiQu95}.
The references for the various potentials are: QCD(BT) \cite{BuTy81},
Power law \cite{Ma80}, Logarithmic \cite{QuRo}, Cornell \cite{Ei80}.}
\label{R0cc}
\end{center}
\end{table}

\subsection{The spectra}

The charmonium radial wave function and its derivatives, for various states, 
are reported and compared with other quark model calculations in table 
\ref{R0cc}.

For $S$ wave mesons and for $V_0$, the wave function at the origin is
essentially constant, except for the state $1S$. For the other potentials and 
for the states reported in table \ref{R0cc}, $R(0)$ decreases when the 
principal quantum number increases. This is a major difference between $V_0$ 
and the other potentials. For $V_0$ this means that the decays widths which 
are sensible to the wave function at the origin, such as the leptonic decays, 
should scale approximately with the inverse of meson mass squared. For the 
other potentials this scaling behaviour is not verified. Looking at $R(0)$, 
the potential which is closer to $V_0$ is the Cornell potential. 

For $P$ mesons, the first derivative of the wave function at the origin
increases with the principal quantum number for the QCD inspired potentials
$V_0$, QCD(BT) and Cornell. The largest increase reported in table \ref{R0cc}
happens for $V_0$. For the other two potentials, $| R^{(1)} (0) |^2$
is essentially constant. 

For the only $D$ meson reported in table \ref{R0cc}, $V_0$ shows a second 
derivative at the origin an order of magnitude larger than all other 
potentials. Thus, $V_0$ enhances, for example, the leptonic decay of 
the $D$ meson states when compared with all other models.

The full theoretical charmonium spectra is reported in table \ref{charmspec},
together with experimental spectra, including the recently discovered 
charmonium states $X(3872)$ \cite{Ab052,Ab053}, $Z(3930)$ 
($\chi_{c2} (2P)$ in the latest version of the Particle Data Book \cite{PDG})
 \cite{Ab05}, 
$X(3940)$ \cite{Ab051} and $Y(4260)$ \cite{Au05}. The experimental values are 
from the Particle Data Book \cite{PDG}, with the exception of the new states. 
The meson masses are defined according to
\begin{equation}
   M( \mbox{meson} ) ~ = ~ 2 m_c \, + \, E_{NR} \, + \, \Delta_{cc} \, ,
\end{equation}
where $E_{NR}$ is the eigenvalue of the Schr\"odinger equation and 
$\Delta_{cc} = -3223$ MeV is a shift introduce to reproduce the 
$J/\Psi$ experimental mass value. This mass shift, $\Delta_{cc}$, can be 
viewed as a non-perturbative contribution which is not accessible within the 
model.

On overall, one can find good agreement between the experimental numbers and 
the model predictions. The deviations reported in table \ref{charmspec}
are below 3 \% level. 

The quantum numbers of various charmonium states have never been measured 
experimentally. In the following, we comment on our results and possible 
quantum number assignments.

\begin{table}[t]
\vspace*{-0.7cm}
\begin{center}
\begin{small}
\begin{tabular}{c||c|c|c|c}\hline\hline
state & Theory & Particle & Experimental & deviation (MeV)    \\
\hline\hline
$1^3S_1$ & 3097$^{\ast}$ & $J/\Psi(1S)$ & $3096.916\pm 0.011$ &  \\
$1^1S_0$ & 3075& $\eta_c$(1S) &$2980.4\pm 1.2$ & 95 \\\hline
$1^3P_2$ & 3556$^{\ast}$ & $\chi_{c2}$(1P)&$3556.26\pm 0.11$ & \\
$1^3P_1$ & 3462  &  $\chi_{c1}$(1P) &$3510.59\pm 0.10$& -49 \\
$1^3P_0$ & 3372  &  $\chi_{c0}$(1P) &$3415.16\pm0.35$& -43 \\
$1^1P_1$ & 3478  &  $h_c(1P)$  & $3525.93\pm 0.27$& -48 \\\hline
$2^3S_1$ & 3659  & $\Psi(2S)$ & $3686.093\pm 0.034$& -27 \\
$2^1S_0$ & 3633  & $\eta'_c(2S)$  & $3638 \pm 4$ & -5\\\hline
$1^3D_3$ & 3841  & $\Psi(3836)$  & $3836\pm 13$& 5 \\
$1^3D_2$ & 3755  & & &  \\
$1^3D_1$ & 3688  & $\Psi(3770)$  &$3771.1\pm 2.4$& 83  \\
$1^1D_2$ & 3754  & & & \\ \hline
$2^3P_2$ &4048   & & & \\
$2^3P_1$ &3938   & $X(3872)$ &$3871.2\pm 0.6$ & 67 \\
$2^3P_0$ &3832   & $Y(3940)$ &$3943\pm 17$ & -111  \\
$2^1P_1$ &3959   & $X(3940)$ &$3943\pm 8$ & 16  \\ \hline
$1^3F_4$ & 4114  & & &  \\    
$1^3F_3$ & 4012  & & & \\
$1^3F_2$ & 3932  & $\chi_{c2}(2P)$ & $3929\pm 5\pm 2$& 3  \\
$1^1F_3$ & 4007  & & &   \\ \hline
$3^3S_1$ & 4164  & $Y(4260)$ & $4260 \pm 10$ & -96\\
$3^1S_0$ & 4135  & & & \\\hline
$2^3D_3$ & 4327  & & & \\
$2^3D_2$ & 4230  & & & \\
$2^3D_1$ & 4155  & $\Psi(4160)$  &$4153\pm 3$ & 2 \\
$2^1D_2$ & 4230  & & & \\
\hline
$1^3G_5$ & 4382 & & & \\
$1^3G_4$ & 4260 & & & \\
$1^3G_3$ & 4161 & & & \\
$1^1G_4$ & 4253 & & & \\
\hline
$3^3P_2$ & 4546 & & & \\
$3^3P_1$ & 4422 & & & \\
$3^3P_0$ & 4300 & & & \\
$3^1P_1$ & 4446 & & & \\
\hline
$2^3F_4$ & 4603 & & & \\
$2^3F_3$ & 4492 & & & \\
$2^3F_2$ & 4406 & & & \\
$2^1F_3$ & 4487 & & & \\
\hline
$4^3S_1$ & 4669 & & & \\
$4^1S_0$ & 4638 & & & \\
\hline
$3^3D_3$ & 4825 & & & \\
$3^3D_2$ & 4719 & & & \\
$3^3D_1$ & 4636 & & & \\
$3^1D_2$ & 4720 & & & \\
\hline
$2^3G_5$ & 4877 & & & \\
$2^3G_4$ & 4747 & & & \\
$2^3G_3$ & 4641 & & & \\
$2^1G_4$ & 4740 & & & \\
\hline\hline
\end{tabular}
\end{small}
\caption{Charmonium spectrum up to $5000$ MeV. The table includes only
         $L \le 4$ states. The quantum states with $^\ast$ where used to set 
         the model parameters ($m_c$, $\Delta_{cc}$).}
\label{charmspec}
\end{center}
\end{table}

\subsubsection{$\psi (3836)$}

The experimental information on $\psi (3836)$ is scarce and this state no 
longer appears in the last particle data book \cite{PDG}. Only two
experiences \cite{Ant94,Bai98} have seen signs for this meson.
The measure of the quantum numbers of $\psi (3836)$ was discussed in 
\cite{Ant94}. Based on the comparisation with theoretical mass predictions, 
the authors suggest that it should be a $^3D_2$ state. Following the same 
reasoning, from the values of the mesons mass and considering the already 
excluded quantum numbers, this state can be identified with a 
$1^3D_3$. The mass difference between the prediction and the 
experimental figures is 5 MeV ($0.1\%$).

\subsubsection{$X (3872)$}

This particle was discovered by Belle collaboration \cite{Choi03} in the decay
$B^- \rightarrow X(3872) K^-$, with $X \rightarrow J / \psi \pi^+ \pi^-$,
and confirmed by CDF \cite{Ac04}, D0 \cite{Ab04} and BABAR \cite{Aub05}
collaborations. Experimentally, only the charge parity $C=+$ is well 
established with the data favouring \cite{Swa05,Qui05,Les05} the assignment 
$J^{PC} = 1^{++}$. A $J^{PC} = 2^{++}$ is not ruled out but seems to be 
unlikely. In our calculation, $1^1D_2$, $2^3P_1$, $1^3F_2$ mesons have masses
close to the mass of $X(3872)$ and with $C=+$. The assignement
with $2^3P_1$ state shows a deviation from the experimental mass of
67 MeV ($1.7\%$).

\subsubsection{The 3940 MeV mass states}

Belle collaboration reported on the three possible new states with masses
around 3940 MeV: $X(3940)$ \cite{Ab051}, $Y(3940)$ \cite{Choi05Y} and 
$Z(3930) / \chi_{c2} (2P) $ \cite{Ab05} with measured masses of 
$3943 \pm 6 \pm 6$ MeV, 
$3940 \pm 11$ MeV and $3931 \pm 4$ MeV, respectively. The $X(3940)$ was 
observed in the process $e^+e^- \rightarrow J / \psi X$, $Y(3940)$ was seen as
a resonance in the decay $B \rightarrow K \pi\pi\pi J / \psi$ and $Z(3930)$ 
was observed in the reaction 
$\gamma \gamma \rightarrow D^0 {\overline D}^0, \, D^- D^+$. For the $Z$
particle, the helicity distribution of the final states particles favors
a $J = 2$ assignment. In the update version of Particle Data Book \cite{PDG} 
it appears as a $\chi_{c2}(2P)$ state with mass value $3929\pm 5\pm 2$.

Looking at the mass values computed with the model under discussion, the
natural candidates are $2P$ and $1^3F_2$ levels. For $Z$, assuming that it 
has $J=2$ and $C=+$, it should be a $1^3F_2$ charmonium state. The 
theoretical computed mass shows a deviation from the experimental number of 
3 MeV ($0.0\%$).

In what concerns the quantum numbers, for the remaining two states there is no
experimental information. Looking at the mass values, the only candidates for
$X(3940)$ and $Y(3940)$ are $2^3P_0$ and $2^1P_1$ states. 
Since the identification of $X$ with a $\chi_{c1}(2P)$ state has problems 
\cite{Swan06}, one assumes that it is a $2^1P_1$ state. The theoretical 
prediction is 111 MeV (deviation of 2.8\%) lower than the experimental value. 
For $Y(3940)$ the mass suggests a $\chi_{c0} (2P)$ state. Such state has
positive parity state, in agreement with the prediction of \cite{Ger06}. Note 
that according to these authors, $X(3940)$ should be one of the following 
states $\chi_{c1} (2P)$, $h_c (2P)$ or $\eta_c (3S)$. Their analysis favors 
the $3^1S_0$ state. For $Y$ the theoretical prediction is 16 MeV (deviation
of 0.4 \%) above the value quoted in the particle data book.

\subsubsection{$Y(4260)$}

This broad resonance was observed in 
$e^+e^- \rightarrow \gamma_{ISR} \pi^+ \pi^- J/ \psi$ with a mass of 4.26 GeV
\cite{Au05}, meaning that it has the quantum numbers of the photon 
$J^{PC} = 1^{--}$. From table \ref{charmspec}, the closest mass state is
the $3^3S_1$, which has a mass 96 MeV (deviation of 2.3 \%) lower than
the measured value.

\subsubsection{$\psi (4040)$, $\psi (4160)$, $\psi (4415)$}

The states $\psi (4040)$, $\psi (4160)$, $\psi (4415)$ were observed in
$e^+ e^- \rightarrow hadrons$ \cite{psis}. In what concerns their quantum 
numbers, there is no experimental information but it is usual to assume that
they are vector particles. In what concerns theoretical predicitions,
there aren't $1^{--}$ states around the mass value of $\Psi(4040)$ and
$\Psi (4415)$. With quantum numbers $1^{--}$ we can assign the $\Psi(4160)$
with $2^3D_1$. 

From table \ref{charmspec} and looking only at the mass values, there a number
of states which can be associated with $\Psi (4040)$ and $\Psi(4415)$. For
the first state, $2^3P_2$ and $1^3F_3$ are good candidates and for
$\Psi(4415)$ the candidates are $3^3P_1$ and $3^1P_1$. Note that, given the 
difference in mass to the lowest $1^{--}$ meson, 942 MeV for $\Psi (4040)$ and
1324 MeV for $\Psi(4415)$, a possible interpretation for both states being 
that they are gluonic excitations rather than quark excitation states. 
In this sense, $\Psi (4040)$ should be a bound state of $J / \Psi$ and the 
$J^{PC}=0^{++}$ $f_0 (980)$, while $\Psi(4415)$ a bound state of $J / \Psi$ 
and the $J^{PC}=0^{++}$ $f_0 (1324)$. Given the lack of experimental 
information, hopefully, only after studying their production and decays will 
be possible to identify the associated quantum numbers.

\subsection{Comparisation With Other Quark Models and With Lattice QCD}

In order to compare our spectra with other charmonium calculations, we 
reproduce table II from \cite{Swan06} including our figures in 
table \ref{QMLV0} and updating the particle data book figures. 
In table \ref{QMLV0}, BGS are the results of a simple non-relativistic quark 
model \cite{BGS}, GI is an updated version of the ``relativised'' 
Godfrey-Isgur model \cite{BGS}, EFG is the relativistic quark model of Ebert, 
Faustov and Galkin \cite{EbFa03}, Cornell are the results of the original 
Cornell model including effects to the coupling to open charm virtual states 
\cite{Ei80}. CP-PACS \cite{Ok02} and Chen \cite{Chen} are two lattice 
calculations in the quenched approximation. For comparisation, in the first 
column we report the Particle Data Book values \cite{PDG}. Note that in table,
the particle identification assumes the ``Standard'' notation, i.e. 
$3^3S_1$ is identified with $\Psi(4040)$,  $4^3S_1$ with $\Psi(4415)$ and
$Z(3930)$ with $\chi_{c2} (2P)$. From the table it is clear that $V_0$ as well
as the lattice simulations underestimate the hiperfine splittings 
$^3S_1 - ^1S_0$. For the lattice this is believed to be due to the quenched
approximation - see, for example, \cite{trinlat05}.

\begin{table}[t]
\begin{center}
\begin{tabular}{c|c|ccccccc}\hline
State & PDG & $V_0$ & BGS  & GI  & EFG  & 
Cornell  & CP-PACS   & Chen \\                                 
\hline\hline
$J/\Psi(1^3S_1)$ & $3096.916 \pm 0.011$ 
        &3097 & 3090 & 3098 & 3096 & 3095 & $3085\pm 1$ & $3084\pm 4$ \\
$\eta_c(1^1S_0)$ & $2980.4 \pm 1.2$ 
        &3075 &2982 &2975 & 2979 & 3095 & $3013\pm 1$ & $3014\pm 4$  \\ 
\hline
$\Psi'(2^3S_1)$ & $3686.093 \pm 0.034$ 
        &3659 & 3672 & 3676 & 3686 & 3684 & $3777\pm 40$ & $3780\pm 43$ \\
$\eta_c(2^1S_0)$ & $3638 \pm 4$ 
        &3633 & 3630 & 3623 & 3588 & 3684 & $3739\pm 46$ & $3707\pm 20$  \\ 
\hline
$\Psi(3^3S_1)$ & $4039 \pm 1$ 
        &4164 & 4072 & 4100 & 4088 & 4225 & -- & --  \\
$\eta_c(3^1S_0)$ & &4135 & 4043 & 4064 & 3991 & 4110 & -- & -- \\ 
\hline
$\Psi(4^3S_1)$ & $4421\pm 4$
        &4669 & 4406 & 4450 & -- & 4625 & -- & --  \\
$\eta_c(4^1S_0)$ & &4638 & 4384 & 4425 &-- & 4460 & -- & -- \\ 
\hline
$\chi_2(1^3P_2)$ & $3556.20 \pm 0.09$ 
        &3556 & 3556 & 3550 & 3556 & 3523 & $3503\pm 24$ & $3488\pm 11$  \\
$\chi_1(1^3P_1)$ & $3510.66 \pm 0.07$ 
        &3462 & 3505 & 3510 &3510 & 3517 & $3472\pm 9$ & $3462\pm 15$ \\
$\chi_0(1^3P_0)$ & $3414.76 \pm 0.35$ 
        &3372 & 3424 & 3445 & 3424 & 3522 & 3408 & $3413\pm 10$ \\
$h_c(1^1P_1)$ & $3525.93 \pm 0.27$
        &3478 & 3516 & 3517 & 3526 & 3519 & $3474\pm 40$ & $3474\pm 20$ \\ 
\hline
$\chi_2(2^3P_2)$ & $3929 \pm 5 \pm 2$  
        &4048 & 3972 & 3979 & 3972 & -- & $4030\pm 180$ & -- \\
$\chi_1(2^3P_1)$ & 
        &3938 &  3925 & 3953 & 3929 & -- & $4067\pm 105$ & $4010\pm 70$\\
$\chi_0(2^3P_0)$ & 
        &3832 &  3852 & 3916 & 3854 & -- & $4008\pm 122$ & $4080\pm 75$\\
$h_c(2^1P_1)$ &    
        &3959 &  3934 & 3956 & 3945 & -- & $4053\pm 95$ & $3886\pm 92$\\ 
\hline
$\chi_2(3^3P_2)$ & &4546 &  4317 & 4337 &  --  & -- & --           &    \\
$\chi_1(3^3P_1)$ & &4422 &  4271 & 4317 &  --  & -- & --           &    \\
$\chi_0(3^3P_0)$ & &4300 &  4202 & 4292 &  --  & -- & --           &    \\
$h_c(3^1P_1)$    & &4446 &  4279 & 4318 &  --  & -- & --           &     \\ 
\hline
$\Psi_3(1^3D_3)$ & &3841 &  3806 & 3849 & 3815 & 3810 & -- & $3822\pm 25$\\
$\Psi_2(1^3D_2)$ & &3755 &  3800 & 3838 & 3811 & 3810 & -- & $3704\pm 33$\\
$\Psi_1(1^3D_1)$ & $3771.1 \pm 2.4$ 
         &3688 & 3785 & 3819 & 3798 & 3755 & -- & -- \\
$\eta_{c2}(1^1D_2)$ &  &3754 & 3799 & 3837 & 3811 & 3810 & -- & $3763\pm 22$\\ \hline
$\Psi_3(2^3D_3)$ &  &4327  & 4167 & 4217 & -- & 4190 & -- & -- \\
$\Psi_2(2^3D_2)$ &  &4230  & 4158 & 4208 & -- & 4190 & -- & --\\
$\Psi_1(2^3D_1)$ &  $4153 \pm 3$ &4155 
         & 4142 & 4194 &-- & 4230 & -- & --   \\
$\eta_{c2}(2^1D_2)$ & &4230 & 4158 & 4208 & -- & 4190 & -- & --\\ \hline
$\chi_4(1^3F_4)$ & &4114 & 4021 & 4095 & -- & -- & -- & -- \\ 
$\chi_3(1^3F_3)$ & &4012 & 4029 & 4097 & -- & -- & -- & $4222\pm 140$\\ 
$\chi_2(1^3F_2)$ & &3932 & 4029 & 4092 & -- & -- & -- & --\\ 
$h_{c3}(1^1F_3)$ & &4007 & 4026 & 4094 & -- & -- & -- & $4224\pm 74$\\ \hline
$\chi_4(2^3F_4)$ & &4603 & 4348 & 4425 & -- & -- & -- & -- \\ 
$\chi_3(2^3F_3)$ & &4492 & 4352 & 4426 & -- & -- & -- & -- \\ 
$\chi_2(2^3F_2)$ & &4406 & 4351 & 4422 & -- & -- & -- & -- \\ 
$h_{c3}(2^1F_3)$ & &4487 & 4350 & 4424 & -- & -- & -- & -- \\ \hline
$\Psi_5(1^3G_5)$ & &4382 & 4214 & 4312 & -- & -- & -- & --  \\
$\Psi_4(1^3G_4)$ & &4260 & 4228 & 4320 & -- & -- & -- & -- \\
$\Psi_3(1^3G_3)$ & &4161 & 4237 & 4323 & -- & -- & -- & -- \\
$\eta_{c4}(1^1G_4)$ & &4253 & 4225 & 4317 & -- & -- & -- & -- \\ \hline\hline
\end{tabular}
\caption{Quark models and lattice charmonium spectra.}
\label{QMLV0}
\end{center}
\end{table}

Following \cite{Swan06}, in table \ref{AveErr} we report the average model 
errors using the values reported in table \ref{QMLV0}. Note that, the particle
identification in table \ref{QMLV0} does not favors our calculation.
Nevertheless, the model predictions are in line with other quark models and
provides an overall agreement with the experimental spectra similar to
lattice calculations. Of course, an analysis of the meson decays is required to
confirm or not the model as a valid model for charmonium.

\begin{table}[t]
\begin{center}
\begin{tabular}{ccccccc}\hline
$V_0$ & BGS & GI & EFG & Cornell & CP-PACS & Chen\\
\hline\hline 
$1.7\%$ & $0.3\%$ &   $0.6\%$ &  $0.4\%$ &  $1.8\%$ &   $1.5\%$ &  $1.4\%$\\
$0.6\%$ & $0.1\%$ &   $0.2\%$ &  $0.2\%$ &  $0.7\%$ &   $0.6\%$ &  $0.5\%$\\ 
\hline\hline   
\end{tabular}
\caption{Average Model Errors. The first line is the average error. The second
line is the square root of the sum of the relative deviations squared, divided
by the number of states.}
\label{AveErr}
\end{center}
\end{table}

\section{Charmonium Decays}

In order to further test the heavy quark potential under discussion, we have
computed leptonic and hadronic decay widths.

\subsection{Leptonic Decays}

For the leptonic decays we follow the van Royen-Weisskopf \cite{RW67} approach
and assume that QCD corrections can be included perturbatively. Then, the
computation of the decay widths requires only the knowledge of the wave 
function at the origin and its derivatives \cite{decaimentos}.

For $n^3S_1$ states, the leptonic width $\Gamma_{e^+ e^-}$ is given by
\begin{equation}
  \Gamma_{e^+ e^-}^{(0)}(^3S_1) ~ = ~ 
  \frac{4 \, e^2_q \, \alpha^2}{M^2_{q{\bar q}}} ~ |R_{nS}(0)|^2
  \label{vanroyen}
\end{equation}
where $M_{q{\bar q}}$ is the mass of the $^3S_1$ state, $e_q$ is the electric 
quark charge (in units of $|e|$), $\alpha$ the fine structure constant and
$R_{nS}(0)$ the radial wave function of the bound state at the origin. 

For $n^3D_1$ states, the leptonic width is \cite{Barnes,Gonzalez}
\begin{equation}
\Gamma_{e^+ e^-}^{(0)}(^3D_1) ~ = ~ 
\frac{25 \, e^2_q \, \alpha^2}{2 m_q^4 M^2_{q{\bar q}}} ~ |R^{(2)} (0)|^2
\end{equation}
where $R^{(2)}_{^3D_1}$ stands for the second derivative of the radial wave 
function at the origin. 

In table \ref{ccee} we report our predictions for the leptonic widths. In order
to avoid the problem of the estimation of QCD corrections, the theoretical
widths were computed relative to the $J/ \Psi$ width. The figures reported
in table \ref{ccee} use the particle data book value for 
\begin{equation}
 \Gamma_{ \left( J / \Psi \rightarrow e^+e^- \right) } ~ = ~
   5.55 \pm 0.14 \pm 0.02 \mbox{  keV}.
\end{equation}

\begin{table}[t]
\begin{center}
\begin{tabular}{ccccc}\hline\hline
    &   $\left| R_{nl}^{(l)} (0) \right|^2$   & 
        $\Gamma_{theo}$                       & 
        assign.                               & 
        $\Gamma_{exp}$ \\
\hline
$2^3S_1$ &1.261 & 2.84 &$\Psi(2S)$   & $2.48\pm 0.06$ \\
$3^3S_1$ &1.283 & 2.26 &               &  \\
         &      & 2.28 &$\Psi(4160)$ & $0.83\pm 0.07$ \\
         &      & 2.16 &$\Psi(4260)$ & -- \\

$4^3S_1$ &1.350 & 1.90 &             &                \\
         &      &      &             & \\
$1^3D_1$ &0.031 & 0.02 & $\Psi(3770)$ & $0.219^{+0.028}_{-0.022}$\\
$2^3D_1$ &0.101 & 0.05 & $\Psi(4160)$ & $0.83\pm 0.07$\\
$3^3D_1$ &0.222 & 0.08 &  & \\
\\
\hline\hline
\end{tabular}
\caption{Theoretical leptonic widths relatively to the experimental value 
$\Gamma_{e^+ e^-}(J/\Psi)$ in keV.  The experimental numbers are from Particle
Data Book. The second colum reports the charmonium squared radial radial wave 
function and its derivatives at the origin in GeV$^{3+l}$. In the widths
assign to a given known particle, in the computation of the width it was used 
the experimental meson mass rather than the theoretical prediction.} 
\label{ccee}
\end{center}
\end{table}

In table \ref{decayscc} our predictions for $n^3S_1$ widths are compared with 
other models. All values where computed relatively to the $J / \Psi$ width and
using th experimental value for the meson masses as reported in the Particle 
Data Book \cite{PDG}. All models give reasonable estimates for the
$\Psi (2S)$ leptonic width but for the remaining states the leptonic width are
clearly overestimated, with $V_0$ providing always the largest figures. 
For $V_0$, $\Psi (4040)$ does not fit into the spectra and therefore the number
reported should not compare well with the experimental figure. 
For $\Psi (4160)$, at least in some models, it is expected a large mixing with 
an S-state which could explain the differences between the theoretical 
prediction and the experimental measure.

\begin{table}[t]
\begin{center}
\begin{tabular}{lrrrrrr}
\hline\hline
      &  \begin{small} $V_0$       \end{small}   
      &  \begin{small} QCD         \end{small}
      &  \begin{small} Power       \end{small}
      &  \begin{small} Log         \end{small}
      &  \begin{small} Cornell     \end{small}
      &  \begin{small} Exp.        \end{small}    \\
      &     
      &  \begin{small} (BT)     \end{small}
      &  \begin{small} Law      \end{small}
      &  
      &  
      &    \\
\hline
  $J/ \psi (1S)$ &        &        &         &         &     &   \\
  $~ \rightarrow l^+l^-$ (keV)
                 & ---  & ---  & ---   & ---  & ---
                 & $5.55 \pm 0.14\pm 0.02$ \\
 \hline
  $\psi (2S)$    &        &        &         &         &     &   \\
  $~ \rightarrow l^+l^-$ (keV)
         & $2.84$  & $2.55$ & $2.19$ & $2.00$ & $2.50$ &
         $2.48 \pm 0.06$ \\
  \hline
  $\psi (4040)$ $\left[ 3^3S_1 \right]$
                 &        &        &         &         &      & \\
  $~ \rightarrow l^+l^-$ (keV)
         & $2.41$  & $1.84$  & 1.34  & 1.15 & 1.77 & 
                   $0.86 \pm 0.07$ \\
\hline
  $\psi (4160)$ $\left[ 3^3D_1 \right]$
                 &        &        &         &         &      & \\
  $~ \rightarrow l^+l^-$ (keV)
                 & $2.28$  & $1.73$  & 1.27  & 1.08  & 1.68 & 
                   $0.83 \pm 0.07$ \\  
\hline\hline
\end{tabular}
\caption{Charmonium leptonic decay widths in keV.}
\label{decayscc}
\end{center}
\end{table}

\subsubsection{Mixing Effects in Charmonium}

For $J^{PC} = 1^{--}$ mesons, the theoretical predictions of the leptonic 
widths computed using the heavy quark potential $V_0$ are reported in table 
\ref{ccee}. For $\psi (2S)$ the model prediction is larger than the 
experimental number by a factor of 15\%. The model overestimates the
widths of $\psi (3770)$ and $\psi (4160)$ by a large factor. 
The deviation can be explained, at least partially, in terms of mixing between 
states. In the quark model under discussion, the tensor interaction is 
responsable for the mixing of the S-wave and D-wave central potential states. 
Indeed, $\psi (2S)$ a $2^3S_1$ state, $\psi (3770)$ a $1^3D_1$ state and 
$Y(4260)$ identified with a $3^3S_1$ state, $\Psi (4160)$ identified with a
$2^3D_1$ have the same spin and the same total spin. Moreover, the theoretical
mass predictions for such states differ by 29 MeV and 9 MeV, respectively; see
table \ref{charmspec}. 

The perturbative calculation of the mixing induced by the spin-dependent 
potential improves only very sligthly the agreement between theory and 
experiment. Therefore, the differences are not due to the mixing between the 
single channel states considered here but mainly to the coupling to open charm
channels. In this article we will not perform a coupled channel analysis. 
Instead, a phenomenologically estimate of the mixing for these 
$J^{PC} = 1^{--}$ states, parametrizing the mixing via a single parameter 
\cite{Ros01} 
\begin{eqnarray}
  & &
  | \psi (3770) \rangle ~ = ~ \cos \theta \, | 1^3D_1 \rangle ~ + ~
                                  \sin \theta \, | 2^3S_1 \rangle ~ ,\label{psi3770} \\
  & &
  | \psi (2S) \rangle   ~ = ~ - \sin \theta \, | 1^3D_1 \rangle ~ + ~
                                  \cos \theta \, | 2^3S_1 \rangle ~ ,\label{psi2s}
\end{eqnarray}
The leptonic widths are given by \cite{Nov78}
\begin{equation}
\Gamma_{e^+e^-} \big( \psi (3770) \big) \, = \,
 \frac{4 \alpha^2 e^2_c}{M^2[ \psi (3770) ]} \,
 \left| \sin \theta  \, R_{2S} (0) ~ + ~
        \frac{5}{2 \sqrt{2} m^2_c} \, \cos \theta \, R^{(2)}_{1D} (0)
 \right|^2 
\end{equation}
and
\begin{equation}
\Gamma_{e^+e^-} \big( \psi (2S) \big) \, = \,
 \frac{4 \alpha^2 e^2_c}{M^2[ \psi (2S) ]} \,
 \left| \cos \theta  \, R_{2S} (0) ~ - ~
        \frac{5}{2 \sqrt{2} m^2_c} \, \sin \theta \, R^{(2)}_{1D} (0)
 \right|^2 
\end{equation}
where $\alpha$ is the fine-structure constant and $e_c$ the charm electric 
charge. Similar expressions hold for $Y(4260)$ and $\Psi (4160)$. 
If the QCD corrections are identical for the two states it cames
\begin{eqnarray}
   \frac{ M^2 [ \psi (3770)  ] \, \Gamma_ {e^+e^-} \big( \psi (3770) \big) }
        { M^2 [ \psi (2S)  ] \, \Gamma_ {e^+e^-} \big( \psi (2S) \big) } ~ = 
    & & 
    \left|
       \frac{ 1.123 \sin \theta \, + \, 0.089 \cos \theta }
            { 1.123 \cos \theta \, - \, 0.089 \sin \theta }  \right|^2 ~ = 
    \nonumber \\
    = & &
    0.092 \pm 0.012 \\
\end{eqnarray}
Fitting these ratios to the experimental values yields the solutions 
$\theta = 12.4^o \pm 1.0^o$, $-21.4^o \pm 1.0^o$ and the leptonic decay widths 
become
\begin{displaymath}
 \begin{array}{lrrr}\hline\hline
             & \Gamma_ {e^+e^-} & \Gamma_ {e^+e^-}   & 
                                  \Gamma_ {e^+e^-}   \\
             & \mbox{no mixing} & \mbox{with mixing} &  \mbox{Exp.} \\ \hline
 \psi (2S)   & 2.84             & 2.62               & 2.48 \pm 0.06 \\
 \psi (3770) & 0.02             & 0.23               & 0.219^{+0.028}_{-0.022}
\\ \hline\hline
 \end{array}
\end{displaymath}
with all widths in keV. Clearly, the mixing improves the theoretical
estimations.

Assuming that $Y(4260)$ is a $3^3S_1$ state which mix with $\Psi (4160)$ a
$2^3D_1$ state, or reversing the assignments, and adjusting the mixing angle
to reproduce the experimental width 
$\Gamma_{e^+ e^-}(\Psi(4160))=0.83\pm 0.07$ keV one obtains
$\Gamma_{e^+ e^-}(Y(4260)) = 1.42$ keV in both cases. This calculation
is resumed in the following table.
\begin{displaymath}
 \begin{array}{lrrrr}\hline\hline
 state & assignment    & \Gamma_ {e^+e^-} & \Gamma_ {e^+e^-}   & 
                                  \Gamma_ {e^+e^-}   \\
       &      & \mbox{no mixing} & \mbox{with mixing} &  \mbox{Exp.} \\ \hline
 3^3S_1 & \Psi (4160)   & 2.28             & 0.83               & 0.83\pm 0.7\\
 2^3D_1 & Y (4260) & 0.04             & 1.42              & \\
\hline
 3^3S_1 & Y (4260)   & 2.16             & 1.42               & \\
 2^3D_1 & \Psi (4160) & 0.05             & 0.83              & 0.83\pm 0.7
\\ \hline\hline
 \end{array}
\end{displaymath}

\subsection{Hadronic widths}

In what concern hadronic decays, for $S$-wave mesons one can consider the 
following processes\footnote{For $n^3S_1$ mesons, one can computed theoretical
estimates for partial widths for decays into $\gamma\gamma\gamma$, $ggg$ 
and $gg\gamma$. However, typically, $^3S_1$ states couple to a large number of
hadronic decay channels, which makes the comparisation of the theoretical 
estimates with experimental figures quite difficult. Moreover, the QCD 
corrections are large and probably the factorisation used in the theoretical 
calculations does not hold. Indeed, typically, the theoretical figures are
quite far away from the experimental values. For these reasons, we do not
discuss hadronic widths for $^3S_1$ mesons.}
\begin{equation}
 n ^1S_0 ~ \longrightarrow  ~  \gamma\gamma, ~ gg \, .
\end{equation}
The theoretical estimates of widths, including QCD corrections, are given by 
(see, for example \cite{KwoMac88})
\begin{eqnarray}
 & &
\Gamma(^1S_0\rightarrow \gamma\gamma) ~ = ~ 3 e_q^4 \, \alpha^2 \,
       \frac{|R(0)|^2}{m^2_Q} \, \left(1-\frac{3.4\alpha_s}{\pi}\right)  \\
 & & 
\Gamma(^1S_0\rightarrow gg) ~ = ~ \frac{2\alpha_s^2}{3} \, 
       \frac{|R(0)|^2}{m^2_Q} \, \left(1+\frac{4.8\alpha_s}{\pi}\right)
\end{eqnarray}
with the replacemente $m_c = M(meson)/2$.

For $n^1S_0$ states, again considering only ratios to the ground state meson,
and using the following experimental values for 
$\Gamma(\eta_c\rightarrow \gamma\gamma)=6.7^{+0.9}_{-0.8}$ keV and  
$\Gamma(\eta_c\rightarrow gg)=\Gamma_{tot}=25.5\pm 3.4$ MeV \cite{PDG}, we 
obtain the widths reported in table \ref{hadS}. For $\eta_c (2S)$ the
theoretical estimated total width agrees well with the measured value.

\begin{table}[t]
\begin{center}
\begin{tabular}{cccc}\hline\hline
Level    &  Final state & Predicted width & Measured width \\
\hline 
$2^1S_0$ & gg              & 12.4 (MeV) & $14\pm 7$ (MeV)  \\
         & $\gamma\gamma$  & 3.26 (keV) & \\
$3^1S_0$ & gg    & 9.77 (MeV) & \\
         & $\gamma\gamma$ &  2.57 (keV)           &                 \\
\hline\hline
\end{tabular}
\caption{$n^1S_0$ widths. In the calculation, for $2^1S_0$ we have used the
experimental mass, while for the $3^1S_0$ we used the predicted mass value.
The measured width reported is the total particle width.}
\label{hadS}
\end{center}
\end{table}  

For $P$ wave mesons the decays we consider the following decays
\begin{eqnarray}
 n ^3P_2 & \longrightarrow &  \gamma\gamma, ~ gg \, ,
 \nonumber \\
 n ^3P_1 & \longrightarrow &  q {\overline q} g \, ,
 \nonumber \\
 n ^3P_0 & \longrightarrow &  \gamma\gamma, ~ gg \, 
 \nonumber \\
  n ^1P_1 & \longrightarrow &  \gamma\gamma, ~ ggg, ~ gg\gamma \, .
 \nonumber \\
\end{eqnarray}
The expressions for the decay widths relevant to $P$ states, without QCD 
corrections, are
\begin{eqnarray}
 & &
\Gamma(^3P_0\rightarrow gg)  ~ = ~ 6 \alpha_s^2 \,
          \frac{|R_{P}^{(1)}(0)|^2}{m^4_Q} \\
 & &
\Gamma(^3P_0\rightarrow \gamma\gamma) ~ = ~ 27 e_q^4 \alpha^2 \,
           \frac{|R_{P}^{(1)}(0)|^2}{m^4_Q} \\
 & &
\Gamma(^3P_1\rightarrow q{\bar q}g) ~ = ~ \frac{8 n_f \alpha_s^3}{9\pi} \,
           \frac{|R_{P}^{(1)}(0)|^2}{m^4_Q} \, \ln(m_Q<r>) \\
 & &
\Gamma(^3P_2\rightarrow gg) ~ = ~ \frac{8 \alpha_s^2}{5} \,
          \frac{|R_{P}^{(1)}(0)|^2}{m^4_Q}\\
 & &
\Gamma(^3P_2\rightarrow \gamma\gamma) ~ = ~ \frac{36}{5} e_q^4 \alpha^2 \,
          \frac{|R_{P}^{(1)}(0)|^2}{m^4_Q}\\
 & &
\Gamma(^1P_1\rightarrow ggg) ~ = ~\frac{20\alpha_s^3}{9\pi} \,
          \frac{|R_{P}^{(1)}(0)|^2}{m^4_Q} \, \ln(m_Q<r>) \\
 & &
\Gamma(^1P_1\rightarrow gg\gamma) ~ = ~ \frac{36}{5} e_q^2 \, 
          \frac{\alpha}{\alpha_s} \, \Gamma(^1P_1\rightarrow ggg)
\label{eqGP}
\end{eqnarray}
where $\left< r \right>=\int_{0}^{\infty} dr r u^2_{nl}(r)$.
The expressions for $nP$ states are from \cite{KwoMac88}, except (\ref{eqGP}) 
which is from \cite{Bar04}. The predicted widths are reported in table 
\ref{had}. In the calculation we have $\alpha_s=0.29\pm 0.03$ 
(see \cite{Seth05}), neglect QCD corrections and used $M(state)/2$ rather than
$m_q$ in the formulas, except in the logarithm, where we used $M(state)<r>$.
Note that these formulas are rough estimates of the partial widths for the 
corresponding annihilations processes.

The rates for annihilations of the $^3D_J$ states into three gluons 
(via color-singlet operators) are given by \cite{Bel87,Kwon88}
\begin{equation}
\Gamma(^3D_J\rightarrow ggg)=\frac{10\alpha_s^3}{9\pi}C_J \frac{\left|R_{D}^{''}(0)\right|^2}{m^6_Q}\ln( 4 m_Q <r>)
\end{equation}
where $C_J=\frac{76}{9}$, 1, 4, for $J = 1$, $2$, $3$. 

The two-gluon annihilation rate of the $^1D_2$ state is given by \cite{Nov78}:
\begin{equation}
\Gamma(^1D_2\rightarrow gg)=\frac{2\alpha_s^2}{3} \frac{\alpha_s^3\left|R_{D}^{''}(0)\right|^2}{m^6_Q}
\end{equation}
The values in table \ref{had} are similar to the ones calculated by 
\cite{Bar04}, except for the state $1^3D_1$ where they used $M(1^3D_1)=3872$
MeV, instead of the experimental value. 

\begin{table}[t]
\begin{center}
\begin{tabular}{ccccc}\hline\hline
Level    & assign. &  Final state & Predicted width & Measured width \\
\hline 
$1^3P_0$ & $\chi_{c0}(1P)$ & gg              & 7.22 (MeV) & $10.4\pm 0.7$ (MeV) (a) \\
         &                 & $\gamma\gamma$  & 4.06 (keV) & \\
$1^3P_1$ & $\chi_{c1}(1P)$ & $q{\bar q}g$    & 86.2 (keV) & $0.89\pm 0.05$ (MeV) (a)\\
$1^3P_2$ & $\chi_{c2}(1P)$ & $gg$            & 1.63 (MeV) & $1.55\pm 0.11$ (MeV) (b) \\
         &                 & $\gamma\gamma$  & 921 (eV)   & $559\pm 83$ (eV) (b) \\
$1^1P_1$ & $h_c(1P)$       &      $ggg$           & 71.5 (keV) & $< 1$ (MeV) (a)\\
         &                 & $gg\gamma$      & 5.76 (keV)\\
$2^3P_0$ & $Y(3940)$       & gg              & 8.27 (MeV) & $87\pm 22\pm 26$ (eV) (a) \\
         &                 & $\gamma\gamma$  & 4.66 (keV) & \\
$2^3P_1$ & $X(3872)$       & $q{\bar q}g$    & 0.103 (MeV) & $< 2.3$ (MeV) (a)\\
$2^3P_2$ &                 & $gg$            & 1.99 (MeV) &  \\
         &                 & $\gamma\gamma$  & 1.12 (keV)   &  \\
$2^1P_1$ & $X(3940)$       & $ggg$           & 84.7 (keV) & \\
         &                 &  $gg\gamma$      & 6.82 (keV)\\\hline
$1^3D_1$ & $\Psi(3770)$    & ggg             & 0.63 (keV) & $23.0\pm 2.7$ (MeV) (a) \\
$1^3D_2$ &                 & ggg             & 0.08 (keV) &  \\
$1^3D_3$ & $\Psi(3836)$    &ggg             & 0.27 (keV) &  \\
$1^1D_2$ &                 & gg             & 0.62 (keV) &  \\
\hline\hline
\end{tabular}
\caption{$P$-wave and $D$-wave widths. 
In the calculation for the meson mass we used
the experimental values of the particle assigned. When no assignment is
given, for the meson mass we use the theoretical prediction.
(a) a total width; (b) see \cite{Seth05} - the value decreases if one includes 
radiative corrections.}
\label{had}
\end{center}
\end{table}

\section{Results and Conclusions}

In this paper it is proposed a generalized Faddeev-Niemi ansatz for the gluon
field - see equations (\ref{AnCX}), (\ref{ansatz2}) and (\ref{ortho}). The
ansatz writes the gluon field in terms of two vector fields and a real
scalar field in the adjoint representation of SU(3). In it simplest 
parameterization, $A_\mu$ is given by (\ref{Asimples}) and the associated 
classical equations of motion become abelian-like equations. Moreover, the 
hamiltonian and spin tensor are the sum of two abelian-like contributions 
from the two vector fields. In Landau gauge, the solutions of the classical 
equations of motion can be writen as plane waves, with a time-like four 
momenta, and with the $A_\mu$ having longitudinal and transverse 
polarizations. A particular class of solutions with zero classical energy
is built. These configurations are asymptoticaly growing exponential functions
in space and in time and are caracterized by a unique mass scale. From these 
solutions, a potential for heavy quark phenomenology is suggested and 
investigated. In the range of 0.2 - 1 fm, the new potential follows 
essentially the singlet potential but differ from it for small and large 
inter-quark separations. The minimization of the square difference between the
potentials gives $\Lambda = 228.0$ MeV.

The potential, including perturbatively the spin dependent contributions, is
then used to investigate charmonium. In what concerns the mass spectra, the 
description is precise up to 3\% level. Typically, the difference between the
theoretical estimation and measured mass is well below de 100 MeV. However,
of the known charmonium states, two mesons do not fit in the model prediction,
namely $\Psi (4040)$ and $\Psi (4415)$. Possible interpretations of these
particles as charmonium states is suggested but not discussed in detail. A
full study of the possibilities rised here for these particles is under way.
Despite of the success of the spectra, the hiperfine splitting cames two low 
in the model. This is seen also in lattice simulations and, in our case,
probably cames from fitting the new potential to the lattice single potential.
We are currently trying to fit the potential parameters to charmonium spectra.
This program is very time consuming but we hope to be able to report the 
results in a reasonable time scale.

A first study of the mesons decays was performed. In what concerns $S$-wave 
leptonic decays, the new potential describes well the widths of the lower 
states. For excited states, the identification is not clear and the 
comparisation with experimental numbers requires further investigations. The
leptonic decays show that, for certain states, mixing is crucial.
We have computed perturbatively the mixing induced by the spin-dependent 
potential. However, the mixing induced by the spin-dependent potential only 
changes slightly our results. On one hand this shows that our calculation is 
consistent. On the other hand shows that a coupled channel analysis is 
required to reproduce the observed behaviour. We are currently starting such
a program.

In what concerns the hadronic decays, the model predictions are in reasonable
agreement with total widths estimates for $^1S_0$ and $P$-wave mesons.

The results for the charmonium spectrum and decays suggest that the potential 
and, hopefully, the classical configurations discussed here can teach us 
something about heavy quarkonium. At least, they seem to provide a starting 
point for a more elaborated calculation, namely a coupled channel calculation,
including possible exotic contributions. We are currently involved in 
exploring such possibilities and extend the use of the potential to other
heavy quarkonium.

\section*{Acknowledgements}

R. A. Coimbra acknowledges financial support from Funda\c{c}\~ao para 
Ci\^encia e Tecnologia, grant BD/8736/2002.


\end{document}